\documentclass[aps,prl,twocolumn,showpacs]{revtex4-1}

\usepackage{pdfpages}
\usepackage{graphicx}
\usepackage{subfigure}
\usepackage[utf8]{inputenc}
\usepackage{amsmath}
\usepackage{amssymb}
\usepackage{hyperref}
\usepackage{color}

\DeclareMathAlphabet{\mathpzc}{OT1}{pzc}{m}{it}

\newcommand{\nc}{\newcommand}
\nc{\bra}[1]{\langle #1|}
\nc{\ket}[1]{|#1\rangle}
\nc{\braket}[1]{\left\langle #1 \right\rangle}
\nc{\la}{\langle}
\nc{\ra}{\rangle}
\nc{\Tr}{\text{Tr}}
\nc{\tr}{\text{tr}}
\nc{\e}{\text{e}}
\nc{\Id}{\mathbb{1}}
\nc{\eps}{\varepsilon}
\nc{\bigO}{\mathcal{O}}
\nc{\dd}{\mathrm{d}}
\nc{\dN}{\mathrm{d}N}
\nc{\half}{\frac{1}{2}}
\nc{\app}[1]{Appendix~\ref{#1}}
\nc{\sdot}{\,\bullet}

\nc{\gentr}{\text{Tr}}

\nc{\asp}[1]{\textcolor{red}{[{\bf SP}: #1]}}
\nc{\alg}[1]{\textcolor{blue}{#1}}

\begin{document}
\renewcommand{\sectionautorefname}{Sect.}
\renewcommand{\subsectionautorefname}{Sect.}
\renewcommand{\figureautorefname}{Fig.}
\def\equationautorefname~#1\null{%
  Eq.~(#1)\null
}

\title{Time-delayed quantum feedback control}

\author{Arne L. Grimsmo}\email{arne.loehre.grimsmo@usherbrooke.ca}
\affiliation{D\'epartement de Physique, Universit\'e de Sherbrooke, Sherbrooke, Qu\'ebec, Canada J1K 2R1}

\date{\today}

\begin{abstract}
A theory of time-delayed coherent quantum feedback is developed. More specifically, we consider a quantum system coupled to a bosonic reservoir creating a unidirectional feedback loop. It is shown that the dynamics can be mapped onto a \emph{fictitious} series of cascaded quantum systems, where the system is driven by past versions of itself. 
The derivation of this model relies on a tensor network representation of the system-reservoir time-propagator. 
For concreteness, this general theory is applied to a driven two-level atom scattering into a coherent feedback loop. We demonstrate how delay effects can qualitatively change the dynamics of the atom, and how quantum control can be implemented in the presence of time-delays. 
\end{abstract}




\maketitle

\emph{Introduction.---}Delayed autonomous feedback, where a signal is directly fed back to a system after a controllable time-delay, is an important control tool for classical systems~\cite{Pyragas92,Ahlborn04,Pyragas06}. It is highly attractive as a tool for stabilizing non-equilibrium states of fast dynamical systems, where avoiding any time-costly signal-processing is crucial. Such stabilization is of great experimental and technological relevance~\cite{Boccaletti00,Fradkov05,Michiels07}. In particular, delayed autonomous feedback has been used to stabilize the high frequency dynamics of optical systems and high speed electrical circuits~\cite{Socolar94,Gauthier94}. 

Autonomous feedback is also receiving substantial and growing interest for controlling quantum systems~\cite{Wiseman94,Lloyd00,James08,Mabuchi08,Gough09,Iida12,Kerckhoff12,Shankar13}. 
Because of the relatively short coherence time and fast dynamics of quantum systems, very 
fast feedback control possible with autonomous feedback is highly desirable. 
In addition, any measurement of the feedback signal will necessarily destroy 
its quantum character, making a fully quantum mechanical feedback loop that 
preserves coherence attractive from a fundamental point of view. 
Compelling evidence that this type of coherent feedback can outperform any measurement-based counterpart for important quantum information processing tasks has been given~\cite{Jacobs14,Yamamoto14}. 

A natural way of implementing coherent feedback control loops is by coupling remote quantum systems via waveguides~\cite{vanLoo13,Roch14,Goban14,Gustafsson14}. Time-delays are unavoidable in practice in such setups and are likely to become important if current experiments are scaled up to larger and more complex networks~\cite{Kimble08,Devoret13,Barends14}. Despite of this, relatively little theoretical research has been done on delay effects for coherent quantum feedback. A major obstacle is the lack of tractable and general theoretical models for treating the highly non-Markovian dynamics induced by this type of feedback.
The theoretical difficulty lies in the quantum correlations between the control target system and the in-loop quantum field: The field cannot simply be traced out, and one has to deal with a highly entangled quantum state over a continuum of degrees of freedom.

Previous investigations have typically been limited to negligible delays~\cite{Gough09b,Kockum14}, linear systems~\cite{Gough08,Yamamoto14}, or systems with special symmetries such as conservation of excitation number~\cite{Carmele13,Fang15}. 
For linear systems, some very promising theoretical demonstrations of the usefulness of delayed autonomous feedback to stabilize quantum systems have been given recently. In~\cite{Carmele13} it was shown how it can be used to stabilize Rabi oscillations of an atom-cavity system in the single-excitation limit, and in \cite{Sven14} how it can enhance entanglement generated in a biexciton cascade in a quantum dot. Another study demonstrated that delayed coherent feedback might be used as a way of controlling the rate of convergence towards a non-equilibrium steady state in many-atom cavity quantum electrodynamics~\cite{Grimsmo14a}. 

In this letter we go beyond linear systems, and develop a general and tractable theoretical model for time-delayed coherent quantum feedback. This opens up research in a largely unexplored regime of quantum feedback control. In particular, it allows for treating the important case of driven, non-linear systems, something which should be of immediate experimental relevance. We consider a generic setup where an arbitrary quantum system is coupled to a bosonic field forming a feedback loop. We show that the system's density matrix can be found by evolving a time-propagator in an extended system space, followed by a generalized partial trace operation. 
The evolution in this larger space is given by a differential equation for a time-propagator in Lindblad form. Interestingly, we can interpret this evolution as an unconventional quantum cascade~\cite{Carmichael93,Gardiner93}, where the system is driven by past versions of itself. 

The derivation of our model uses so-called tensor network representations of quantum mechanical states and operators~\cite{Orus14}. These tools have their origin at the intersection of condensed matter and quantum information, where they are used to efficiently handle entangled many-body quantum systems. Recently, an intimate connection was made between continuum limits of certain tensor networks and output fields of open quantum systems~\cite{Verstraete10,Osborne10,Gough14}. We develop these ideas further and find a novel application of tensor networks in handling the dynamics of a highly non-Markovian open quantum system.
These developments could be of interest in themselves as a new approach to non-Markovian open systems theory. 

Below, we introduce the model putting emphasis on developing an intuitive picture of the dynamics. Technical details are left to the Supplemental Material~\cite{SM}. As a concrete example we consider a two-level atom coupled to a coherent feedback loop. 
We demonstrate two simple yet remarkable possibilities for delayed feedback control for this example: 1)~spontaneous decay acting only for a controllable time, $\tau$, and 2)~stabilizing Rabi oscillations far beyond the atoms coherence time in the absence of feedback. We discuss how these effects can be observed in a circuit quantum electrodynamics architecture~\cite{Blais04}.

\begin{figure}
    \centering
    \includegraphics{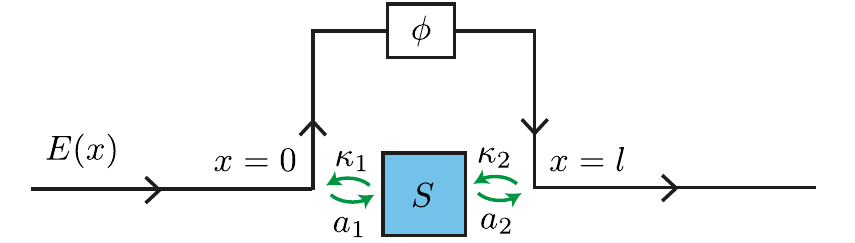}
  \caption{\label{fig:setup}(Color online.) Schematic of the setup. A unidirectional bosonic field, $E(x)$, interacts with the system, $S$, at positions $x=0$ and $x=l$. The interaction at $x=0$ ($x=l$) is with a system operator $a_1$ ($a_2$) and a rate $\kappa_1$ ($\kappa_2$). We assume that an arbitrary phase shift, $\phi$, can be applied to the field between $x=0$ and $x=l$, such that the time-delay, $\tau=l/c$, and the phase are independently controllable.}
\end{figure}

\emph{Physical setup.---}We consider a quantum system coupled to a single unidirectional bosonic field at two different spatial positions, $x=0$ and $x=l$, as depicted in \autoref{fig:setup}. The field mediates a feedback loop for the system~\footnote{We consider a unidirectional field for simplicity. A one-dimensional waveguide with ``left'' and ``right'' moving modes correspond to two independent feedback loops. With equal speed of light in both directions, the situation however maps again to a single loop.}. 
We further assume that an arbitrary phase shift, $\phi$, can be applied to the field between these two positions, such that the time-delay and phase are independent parameters. The system-field Hamiltonian is $H = H_S + H_B + V$, where $H_S$ is the system Hamiltonian, $H_B = \int_0^\infty \dd\omega \omega b^\dagger(\omega) b(\omega)$ the free field Hamiltonian, and $V$ the interaction Hamiltonian,
\begin{align}\label{eq:V}
\begin{aligned}
V =& i \int_{-\infty}^\infty \dd\omega \sqrt{\frac{\kappa_1}{2\pi}} \left(a_1 b^\dagger(\omega) - \text{H.c.}\right)\\
    &+i \int_{-\infty}^\infty \dd\omega \sqrt{\frac{\kappa_2}{2\pi}} \left(a_2b^\dagger(\omega)\e^{-i\omega \tau + i\phi} - \text{H.c.}\right),
\end{aligned}
\end{align}
where $\tau=l/c$ is the time-delay ($c$ the speed of light), $\sqrt{\kappa_{1,2}}$ is the coupling strength at the two positions, $x=0,l$, respectively, $a_1$ and $a_2$ are two system operators, and H.c. stands for Hermitian conjugate. The field modes, $b(\omega)$, satisfy $[b(\omega),b^\dagger(\omega')]=\delta(\omega-\omega')$. For generality, we allow the two system operators, $a_1$ and $a_2$, to be different, but they could very well refer to the same operator---for example the dipole operator of a two-level atom or a cavity mode annihilation operator. The assumptions behind \autoref{eq:V} are standard for open quantum systems, typically valid when the system is described by some frequency $\omega_S \gg \kappa_{1,2}$, see, \emph{e.g.}, Ref.~\cite{Gardiner04}.

To make the discussion more concrete, let us pause to consider a relevant example. A possible implementation is an optical cavity consisting of two mirrors, where the reflected field of one mirror is guided to be used as an input field on the other mirror (the inputs and outputs could be separated by circulators). In this case, one has the interaction in \autoref{eq:V} with $a_1=a_2=a$, for a system annihilation operator $a$, satisfying $[a,a^\dagger]=1$. $\kappa_{1,2}$ are in this example the linewidths of the two respective mirrors. $H_S$ describes the internal dynamics of the cavity, which could be non-linear due to the presence of other quantum degrees of freedom interacting with the cavity field. The equation of motion for the annihilation operator in the Heisenberg picture can be found to be (see, \emph{e.g.}, Ref.~\cite{Gardiner04}):
\begin{align}\label{eq:adot}
\begin{aligned}
\dot{a}(t) =& i[H_S,a(t)] - \half \left(\kappa_1+\kappa_2\right) a(t) - \sqrt{\kappa_1} b_\text{in}(t)\\
&- \sqrt{\kappa_2} \e^{i\phi} \left[b_\text{in}(t-\tau) + \sqrt{\kappa_1}a(t-\tau)\right].
\end{aligned}
\end{align}
Here we have defined an input field $b_\text{in}(t) = \frac{1}{\sqrt{2\pi}} \int_{-\infty}^{\infty} \dd\omega \e^{-i\omega(t-t_0)} b_0(w)$,
where $b_0(\omega)$ are the initial values for $b(\omega)$ in the Heisenberg picture. \autoref{eq:adot} has the form of a delay differential equation~\cite{Kuang93}, and makes the effect of the feedback quite clear. However, since the Heisenberg equations involve coupling between system and field operators, they are typically not efficiently solvable in practice. Also, no corresponding master equation for the reduced system density matrix exists, in general, due to the finite time-delay. In the following, we present a practical scheme to integrate this type of dynamics by embedding the system in a larger space.

\emph{A cascade of information from the past.---}Our main result is a tractable model for the system dynamics after eliminating the field degrees of freedom. The model suggests an intuitive picture where the system is driven by past versions of itself in a cascaded fashion. We here present the model and develop this picture, while leaving the technical details of the derivation to the Supplemental Material~\cite{SM}.

To find the system state, $\rho_S(t)$, at an arbitrary time $(k-1)\tau \le t < k\tau$, for $k=1,2,\dots$, we evolve a time-propagator for a fictitious \emph{cascade} of $k$ identical copies of the system. 
The time-propagator, which we label $\mathcal{E}_s(t)$, is a superoperator on an extended system, $S^{\otimes k}$, and obeys a differential equation in the form of a cascaded master equation as introduced by Carmichael and Gardiner~\cite{Carmichael93,Gardiner93}. Note that we are here considering the master equation as a differential equation for the \emph{propagator}, and not for a density matrix. For simplicity, we consider an incoming vacuum field, and we assume that the system and field are in a product state at time $t=0$ (the in-loop field is also initially in the vacuum state). As shown in the Supplemental Material~\cite{SM}, the differential equation for the propagator then takes the form
\begin{align}\label{eq:cascade}
  \frac{\dd}{\dd s} \mathcal{E}_s(t) = \sum_{l=0}^k \left\{ {-}\frac{i}{2} \mathcal{H}\left[H_{l,l+1}(s)\right] + \mathcal{D}\left[L_{l,l+1}(s)\right] \right\} \mathcal{E}_s(t).
\end{align}
The integration variable, $s$, is an auxilliary time-variable, and the equation is to be integrated up to $s=\tau$, with the initial condition $\mathcal{E}_0(t) \equiv \mathcal{I}^{\otimes k}$, where $\mathcal{I}$ is the system identity super-operator. We have labeled $k$ identical system copies by $S_l$, $l=1,\dots,k$. The superoperators $\mathcal{H}$ and $\mathcal{D}$ are defined by
\begin{align}
    \mathcal{H}[X]\sdot =& [X,\sdot],\\
    \mathcal{D}[X]\sdot =& X\sdot X^\dagger - \half X^\dagger X\sdot - \half \sdot X^\dagger X,
\end{align}
and the operators $H_{l,l+1}$ and $L_{l,l+1}$ are given by
\begin{align}\label{eq:cascops1}
  H_{l,l+1} {=}& H_S^{(l)}{+}H_S^{(l+1)}{+}i\sqrt{\kappa_1\kappa_2}(\e^{i\phi}a_1^{(l)\dagger}a_2^{(l+1)} {-} \text{H.c.}),\\
  L_{l,l+1} {=}& \sqrt{\kappa_1}a_1^{(l)} + \sqrt{\kappa_2}\e^{i\phi}a_2^{(l+1)},
\end{align}
except for $H_{0,1} = H_S^{(1)}$, $H_{k,k+1} = H_S^{(k)}$, $L_{0,1} = \sqrt{\kappa_2} \e^{i\phi}a_2^{(1)}$ and $L_{k,k+1} = \sqrt{\kappa_1} a_1^{(k)}$. 
The superscript denotes the system on which an operator acts. Finally, we have defined $A^{(l)}(s) = A^{(l)}$ for all $l < k$, and $A^{(k)}(s) = \theta[t-(k-1)\tau-s]A^{(k)}$, 
where $\theta(s)$ is the Heaviside step function, for any system operator $A$.
\begin{figure}
    \centering
    \includegraphics{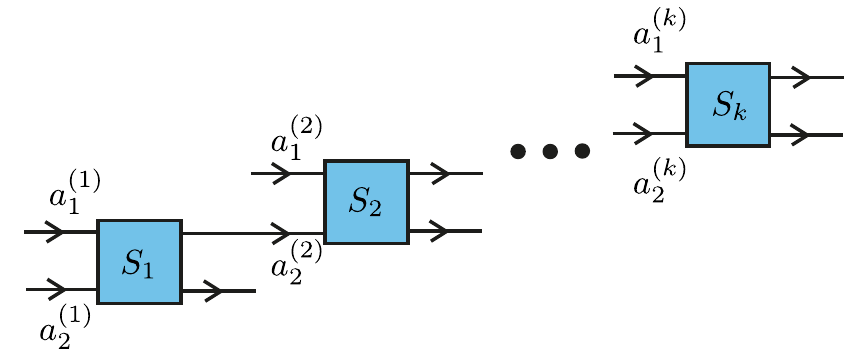}
    \caption{\label{fig:cascade}(Color online.) The time-propagator, $\mathcal{E}_s(t)$, in \autoref{eq:cascade} can be recognized as the propagator for a cascade of $k$ identical systems, $S_l$. We can think of the copies as representing past versions of the system, \emph{i.e.}, the system is being driven by itself from the past.}
\end{figure}

The generator in \autoref{eq:cascade} is exactly the generator for a cascaded chain of $k$ identical quantum systems, as introduced by Carmichael and Gardiner~\cite{Carmichael93,Gardiner93}. An illustration is given in \autoref{fig:cascade}. The evolution would describe a cascade in the usual sense if the time-propagator, $\mathcal{E}_s(t)$, is applied to an initial state on the $k$-fold system space, $S^{\otimes k}$. However, the feedback problem is different, and the solution, $\rho_S(t)$, is found by imposing a peculiar type of ``boundary conditions'' on the propagator, as we will now explain. 

First of all, the integration variable, $s$, in \autoref{eq:cascade} is to be understood as a fictitious time-variable, and the equation is to be integrated up to $s=\tau$, as already stated. $\rho_S(t)$ is then found by acting with $\mathcal{E}_\tau(t)$ on an initial state $\rho_{S_1}(0)$ for the first system, $S_1$, and essentially mapping the output of system $S_l$ to the input of system $S_{l+1}$, for $l=1,\dots,k-1$. The desired solution will be given as the output of system $k$, $\rho_S(t) = \rho_{S_k}(t)$. To explain this in more detail, we first have to introduce a generalized trace operation on the level of superoperators. For a superoperator, $\mathcal{A}$, that acts on a tensor product of \emph{identical} systems, $S_1 \otimes \dots \otimes S_k$, we define the following generalized trace:
\begin{align}
  \gentr_{(S_{l'},S_{l})}\,\mathcal{A}\sdot &= \sum_{ij} \bra{i_l} \mathcal{A}\Big(\sdot\otimes\ket{i_{l'}}\bra{j_{l'}}\Big) \ket{j_l},
\end{align}
where $\ket{i_l}$ and $\ket{i_{l'}}$ are bases for the two respective systems, $S_l$ and $S_{l'}$. Note that with $l=l'$ this is a partial trace, in the usual sense, but on the level of superoperators. More generally, this operation can be understood as mapping the output of system $S_l$ to the input of system $S_{l'}$. 

We are now ready to write down an expression for $\rho_S(t)$, given $\mathcal{E}_\tau(t)$ found from \autoref{eq:cascade}:
\begin{align}\label{eq:rho_S}
    \rho_S(t) = \gentr_{(S_{k},S_{k-1})} \dots \gentr_{(S_2,S_1)} \, \mathcal{E}_\tau(t) \rho_{S_1}(0).
\end{align}
This equation, together with \autoref{eq:cascade}, constitute our main result, as they provide a practical scheme to find $\rho_S(t)$ for an arbitrary time $t$. In practice, the solution is thus found by first integrating \autoref{eq:cascade} up to time $s=\tau$, and subsequently computing the reduced state, $\rho_S(t)$, by acting on the initial state and taking the generalized partial trace in \autoref{eq:rho_S}. To help build an understanding of \autoref{eq:rho_S}, we illustrate the trace operation diagramatically for the case $k=3$ in \autoref{fig:gentrace}. In the Supplemental Material, we give a derivation of Eqs. \eqref{eq:cascade} and \eqref{eq:rho_S} using a tensor network representation of the time-propagator.

\begin{figure}
    \includegraphics{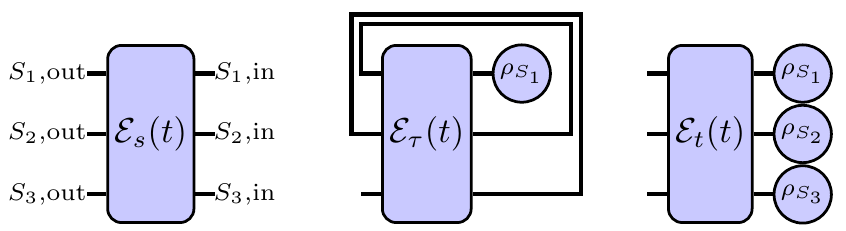}
    \caption{\label{fig:gentrace}(Color online.) A diagrammatic illustration of \autoref{eq:rho_S}. Left: The propagator, $\mathcal{E}_s(t)$, for the case $k=3$. The map is represented by a shape with lines attached to represent the input and output spaces. The labels indicate the systems associated to the lines. Middle: \autoref{eq:rho_S} takes a state as input to system $S_1$, while the output of system $S_1$ is mapped to the input of $S_2$, and similarly the output of $S_2$ to the input of $S_3$. The final output is a state for system $S_3$. Right: For comparison, we show the application of the propagator to a state, $\rho_{S_1}\otimes\rho_{S_2}\otimes\rho_{S_3}$, on the three-fold system space. This case corresponds to a conventional quantum cascade of three identical systems~\cite{Carmichael93,Gardiner93} (the choice of a product initial state is not essential). This diagrammatic notation is developed further in the Supplemental Material~\cite{SM}.}
\end{figure}

How can we now understand the dynamics induced by the feedback field? \autoref{eq:cascade} 
suggest that the dynamics is given by a cascade of instances of the system, where each instance is driven by a past version of itself, from a time $\tau$ earlier. 
What is highly non-trivial is that the feedback field that returns after a time $\tau$, is already quantum correlated with the system it is driving. This leads us to \autoref{eq:rho_S}: it is this equation that correctly account for the quantum correlations \emph{in time} in the cascade picture.

\emph{Delayed coherent feedback for a two-level atom.---}We illustrate the theory with a simple example: a two-level atom coupled to a coherent feedback loop. Both spontaneous decay and resonance fluorescence through the feedback loop is considered.  This setup can, \emph{e.g.}, desribe an atom placed a (large) distance from a mirror, a problem with a long history in quantum optics (see Ref. \cite{Dorner02} and references therein). In the absence of a drive, the problem can be solved analytically due to there being only a single conserved excitation between the system and the reservoir \cite{Milonni74}. In the driven case, the problem has to the best of our knowledge previously only been considered in an approximate sense, employing perturbation theory in various limits \cite{Dorner02}.

The problem is defined by a system Hamiltonian $H_S = \mathcal{E}(\sigma_+ + \sigma_-)$, and coupling operators $a_1=a_2=\sigma_-$. Here $\sigma_- = \ket{g}\bra{e}$ is the atomic lowering operator, and $\sigma_+=(\sigma_-)^\dagger$. $\mathcal{E}$ is the drive amplitude, and we assume that the atom is driven on resonance for $\mathcal{E}>0$. We take the rates to be identical, $\kappa_1=\kappa_2=\gamma$, and assume a phase shift of $\phi=\pi$ in the feedback loop.

\begin{figure}
    \centering
    \includegraphics{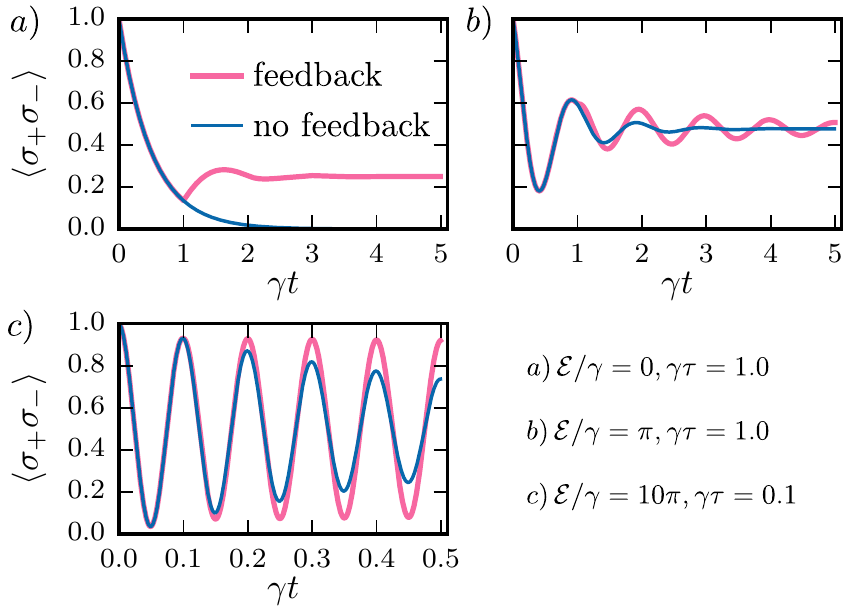}
    \caption{\label{fig:simulation}(Color online.) Time-delayed coherent feedback control of a two-level atom for three different parameter sets, as indicated in the figure. The pink (light gray) lines show the numerical results with feedback, while for comparison the blue (dark gray) lines show analogous simulations without feedback (\emph{i.e.}, $\kappa_1=2\gamma$, $\kappa_2=0$).}
\end{figure}

Numerical results for the solution of Eqs. \eqref{eq:cascade} and \eqref{eq:rho_S} are shown in \autoref{fig:simulation}. The panels show three different cases: $a)$ $\mathcal{E}/\gamma =0$, $b)$ $\mathcal{E}/\gamma = \pi$ and $c)$ $\mathcal{E}/\gamma = 10\pi$. The delay is chosen to be $\gamma\tau = 1.0$ for the case $\mathcal{E}/\gamma = 0$, and otherwise equal to the Rabi oscillation period: $\tau = 2\pi/2\mathcal{E}$. The pink (light gray) lines show results with feedback, while the blue (dark gray) lines are analogous simulations without feedback, for comparison.

We note two remarkable features in \autoref{fig:simulation}: First we consider the simplest case of spontaneous emission in panel $a$. In this case the atom decays exponentially to the ground state in the absence of feedback. In the presence of feedback, however, the feedback field starts driving the system after an initial transient period of time $\tau$, after which the population grows and eventually stabilizes at a steady state value. In steady state, destructive interference between two contributions to the output field, one coming from direct scattering and one from scattering via the feedback loop, prohibits the system from decaying. Hence, we have the possibility of letting the atom decay only for a controllable time. In steady state the system is \emph{dynamically} decoupled from the decay channel. This phenomenon of feedback-induced dynamical decoupling of an atom from a decay channel has been demonstrated previously \cite{Dorner02,Tufarelli14,Whalen15}.

Let us now look at non-zero drive strengths as shown in panels $b$ and $c$ of \autoref{fig:simulation}. Here, the feedback induces long-lived Rabi oscillations, far beyond the coherence time of the atom in the absence of feedback. We have chosen $\tau$ to coincide with the Rabi period, which is an optimal choice for stabilizing the Rabi oscillations. This means that $\tau$ should be considered as a control parameter in its own right. 
In the bottom panel with $\mathcal{E}/\gamma = 10\pi$ and $\tau = 0.1$, the decay is extremely slow after the initial transient period of $\tau$. 
Numerical results have been verified with a brute force numerical integration of the full system plus reservoir dynamics for small values of $\gamma\tau$ \cite{Whalen15b}. This was done by representing the feedback reservoir by a finite number of modes, truncated to have a small total photon number. Such an approach however quickly becomes impractical for large $\gamma\tau$ ($\gtrsim 0.1$).

The simple example we have considered here could be realized experimentally in a variety of different platforms. A particularly appealing implementation is a circuit quantum electrodynamics architecture with an artificial atom coupled to a one-dimensional waveguide~\cite{Blais04,Astafiev10,vanLoo13,Hoi14}.
A meandering waveguide can be made to couple to the artificial atom at two different locations, or the artificial atom can be placed in a semi-infinite waveguide where the endpoint serves as a mirror. Such a setup was recently demonstrated experimentally in~\cite{Hoi14}. A requirement to observe strong delay effects is $\gamma\tau \gtrsim 0.1$, which is readily achievable. In fact, significant delay effects are likely to be unavoidable even for moderate distances for strong coupling between the artificial atom and waveguide.

\emph{Conclusions.---}We have shown that the problem of an arbitrary quantum system coupled to a coherent, field-mediated feedback loop, can be mapped onto a tractable problem in a larger system space. 
This theory also yields an intuitive picture that helps to understand feedback mediated by a quantum field. For practical numerical integration, the approach presented here is superior to alternative approaches based on approximating the feedback reservoir by a lower-dimensional system when the time-delay becomes comparable to the inverse linewidth of the emitting quantum system.

\begin{acknowledgments}
The author thanks Howard Carmichael and Simon Whalen for many useful discussions and for performing a brute force numerical integration to verify the numerical results for small time-delays. The author thanks Alexandre Blais for helpful discussions. This work was supported by NSERC.
\end{acknowledgments}

\bibliography{refs}

\clearpage


\section*{Supplementary material}

# Supplemental Material for "Time-delayed quantum feedback control"

Arne L. Grimsmo*

Département de Physique, Université de Sherbrooke, Sherbrooke, Québec, Canada J1K 2R1
(Dated: July 27, 2015)

This supplemental material to the paper "Time-delayed quantum feedback control" is organized as follows: First we give an introduction to a graphical "tensor network" notation for linear maps, based on Penrose's tensor notation [1, 2], that is central to deriving the main results. We then briefly outline our approach to open quantum systems in a general context. Before approaching the feedback problem of the main letter, we consider the simpler problem of a system coupled to a conventional Markovian reservoir. We use this example to introduce the connection between open system dynamics and continuum limits of certain simple tensor networks. Finally, we consider the feedback problem of the main letter, and derive the main results stated there.

## I. A DIAGRAMMATIC DIRAC NOTATION

In this section we introduce a graphical notation, which we can think of as a diagrammatic Dirac notation. Linear maps, such as vectors and operators, are denoted by shapes with "legs" attached to represent their input and output spaces. This type of notation was also developed at length in the context of open quantum systems in Ref. [2], and we refer to this work for more background information, although the focus there was on finite dimensional vector spaces. We here use slightly different conventions that are more convenient for our purposes.

Let us first introduce some general notation. The Hilbert space of a quantum system $A$ is denoted $\mathcal{H}_A$, where the subscript is used to differentiate between systems whenever there are more than one. Since we are dealing with open quantum systems, the state space of system $A$ is a a subset of the Hilbert space of linear operators acting on $\mathcal{H}_A$. We denote the latter $L(\mathcal{H}_A)$, and refer to it as Liouville space. We can think of a state, $\rho$, either as an operator on $\mathcal{H}_A$, or as a vector in Liouville space $L(\mathcal{H}_A)$. To specify that a map, $\tau$, is an operator on $\mathcal{H}_A$ we write $\tau : \mathcal{H}_A \to \mathcal{H}_A$. To specify that it is a vector in Liouville space we write $\tau \in L(\mathcal{H}_A)$. It should otherwise be clear from context whether an object is thought of as an operator or as a vector. We therefore do not introduce any extra notation for the purpose of differentiating between an operator and the "vectorization" of that operator [12]. We also need to consider linear operators on $L(\mathcal{H}_A)$, which we refer to as superoperators.

We first introduce the diagrammatic notation for an arbitrary complex separable Hilbert space, $\mathcal{V}$, before specializing to the case $\mathcal{V} = L(\mathcal{H}_A)$. A vector $v \in \mathcal{V}$ we denote by a shape with a leg going to the left:

$$v = -\!\!\boxed{v}\, \in \mathcal{V}. \tag{1}$$

Similarly, a vector in the dual space $w^* \in \mathcal{V}^*$, we denote by a shape with the leg going to the right:

$$w^* = \boxed{w}\!\!- \in \mathcal{V}^*. \tag{2}$$

Recall that the dual space, $\mathcal{V}^*$, is the set of all continuous linear functions, $\mathcal{V} \to \mathbb{C}$, and can be identified with $\mathcal{V}$. An operator $A$ on $\mathcal{V}$ we denote by a shape with two legs:

$$A = -\!\!\boxed{A}\!\!- \,:\, \mathcal{V} \to \mathcal{V}, \tag{3}$$

Finally, a number is denoted by a shape without any legs:

$$c = \boxed{c}\, \in \mathbb{C}. \tag{4}$$

---

*Electronic address: arne.loehre.grimsmo@usherbrooke.ca

We can equivalently think of the shapes as tensors, where each leg represents a tensor index, if we allow for countably infinite ranges for the indices [13].

The innerproduct on $\mathcal{V}$, $\langle w, v \rangle$, is denoted

$$\langle w, v \rangle = \underset{w}{\bigcirc}\!\!-\!\!\underset{v}{\bigcirc} \in \mathbb{C}. \tag{5}$$

In general, composing objects is denoted by joining legs. Acting on a vector, $v \in \mathcal{V}$, with an operator, $A$, we denote

$$Av = -\boxed{A}\!-\!\underset{v}{\bigcirc}. \tag{6}$$

Acting on a dual vector would be denoted similarly, with the dual, say $w^*$, placed to the left of $A$, and joining the two nearest legs. In tensor language, joining two legs corresponds to summing over a repeated index [2].

Tensor products of vectors, $w \otimes v \in \mathcal{W} \otimes \mathcal{V}$, are denoted by drawing the vectors vertically aligned, with one above the other:

$$w \otimes v = \begin{matrix} -\underset{v}{\bigcirc} \\ -\underset{w}{\bigcirc} \end{matrix}. \tag{7}$$

An arbitrary vector in the composite vector space, $u \in \mathcal{W} \otimes \mathcal{V}$, is thus naturally drawn

$$u = \begin{matrix} \mathcal{V} \\ \mathcal{W} \end{matrix}\!-\!\boxed{u}, \tag{8}$$

and an operator, $\mathcal{C}$, on $\mathcal{W} \otimes \mathcal{V}$:

$$\mathcal{C} = \begin{matrix} \mathcal{V} \\ \mathcal{W} \end{matrix}\!-\!\boxed{\mathcal{A}}\!-\!\begin{matrix} \mathcal{V} \\ \mathcal{W} \end{matrix} : \mathcal{W} \otimes \mathcal{V} \to \mathcal{W} \otimes \mathcal{V}. \tag{9}$$

In Eq. (8) and Eq. (9) labels are used to indicate the system assoicated to the legs. Such labes are typically omitted whenever they are clear from context.

Finally, it is very useful to introduce a generalized trace operation for operators on tensor product spaces that allows us to connect an "output leg" to an "input leg." Of course, connecting legs only makes sense if the systems associated to the legs are isomorphic. Consider the operator $\mathcal{C}$ in Eq. (9), and assume that $\mathcal{V}$ and $\mathcal{W}$ are isomorphic. We then introduce the following operation

$$\begin{aligned}\text{Tr}_{(\mathcal{Y},\mathcal{X})}\mathcal{C} &: \mathcal{X} \to \mathcal{Y} \\ &= \sum_i x_i^* \mathcal{C} y_i, \end{aligned} \tag{10}$$

where $\mathcal{X}$ and $\mathcal{Y}$ are any of the two spaces $\mathcal{V}$ or $\mathcal{W}$, respectively. $x_i$ and $y_i$ are orthonormal bases for the two respective spaces, $\mathcal{X}$ and $\mathcal{Y}$. We note that this is a partial trace in the usual sense if $\mathcal{X} = \mathcal{Y}$. We denote this operation diagramatically, by connecting the output leg for system $\mathcal{X}$ with the input leg for system $\mathcal{Y}$. For example:

$$\begin{aligned}\text{Tr}_{(\mathcal{W},\mathcal{V})}\mathcal{C} &: \mathcal{V} \to \mathcal{W} \\ &= \boxed{\mathcal{C}} \end{aligned} \tag{11}$$

In tensor language, this operation is of course nothing but summing over the corresponding repeated index.

So far we have used a convention where objects are composed horizontally, right to left, and tensored vertically. We could equivalently have chosen to compose vertically, say top to bottom, and tensor horizontally. Below we take advantage of this freedom, and use different conventions for system and bath objects.

## A. Density operators, superoperators and system-bath tensor products

Let us now consider the case $\mathcal{V} = L(\mathcal{H}_A)$, for a quantum system $A$. Recall that the innerproduct on this space is given by $\langle \sigma_A, \tau_A \rangle = \text{tr}[\sigma_A^\dagger \tau_A]$. We identify the operator-adjoint, $\sigma_A^\dagger$, with the dual vector $\sigma_A^* \in L(\mathcal{H}_A)^*$ of $\sigma_A \in L(\mathcal{H}_A)$.

It is convenient to introduce a special notation for the identity operator, $I_A$, on the system Hilbert space $\mathcal{H}_A$, viewed as a vector in Liouville space:

$$I_A = \!-\!\!\circ \;\in L(\mathcal{H}_A), \tag{12}$$

and its dual

$$I_A^* = \circ\!\!- \;\in L(\mathcal{H}_A)^*. \tag{13}$$

This allows us to draw for the trace of $\tau_A \in L(\mathcal{H}_A)$:

$$\text{tr}\,\tau_A = \langle I_A, \tau_A \rangle = \circ\!\!-\!\!\boxed{\tau_A}. \tag{14}$$

Later, when dealing with system-bath dynamics, it will be useful to have a notation that distinguishes between system and bath objects. We now introduce a set of specialized conventions for this purpose. Consider a system-bath tensor product, $S \otimes B$. System states are denoted as in Eq. (1):

$$\rho_S = -\!\!\boxed{\rho_S} \;\in L(\mathcal{H}_S). \tag{15}$$

We distinguish this from states of the bath, $B$, by drawing the latter with the leg going downwards:

$$\sigma_B = \boxed{\sigma_B} \;\in L(\mathcal{H}_B), \tag{16}$$

A superoperator, $\mathcal{A}_{SB}$, that acts on the composite system-bath space we then draw

$$\mathcal{A}_{SB} = \;\; S \!-\!\boxed{\mathcal{A}_{SB}}\!-\! S \;\;, \tag{17}$$

where the labels indicate the system associated to the legs. This convention is practical for dealing with superoperators acting on product states, $\rho_S \otimes \sigma_B$, which we can now draw

$$\mathcal{A}_{SB}\,\rho_S \otimes \sigma_B = \;\; -\!\boxed{\mathcal{A}_{SB}}\!-\!\boxed{\rho_S}. \tag{18}$$

We need to compose such superoperators in two different ways. Superoperators, say $\mathcal{A}_{SB}$ and $\mathcal{B}_{SB'}$, that act on two different bath systems, $B$ and $B'$:

$$\mathcal{B}_{SB'}\mathcal{A}_{SB} = \;\; S \!-\!\boxed{\mathcal{B}_{S,B'}}\!-\!\boxed{\mathcal{A}_{S,B}}\!-\! S \;\;. \tag{19}$$

And, superoperators that act on the same bath system, $B$, but two different systems $S$ and $S'$:

$$\mathcal{B}_{S'B}\mathcal{A}_{SB} = \;\; \begin{array}{c} S \!-\!\boxed{\mathcal{A}_{S,B}}\!-\! S \\ \\ S' \!-\!\boxed{\mathcal{B}_{S',B}}\!-\! S' \end{array} \;\;. \tag{20}$$

Finally, we consider how the generalized trace defined in Eq. (10) acts on the level of superoperators. Consider a superoperator, $\mathcal{A}$, acting on two isomorphic systems $S$ and $S'$:

$$\mathcal{A}: L(\mathcal{H}_S) \otimes L(\mathcal{H}_{S'}) \to L(\mathcal{H}_S) \otimes L(\mathcal{H}_{S'})$$

$$= \begin{array}{c} S \\ S' \end{array} \boxed{\mathcal{A}} \begin{array}{c} S \\ S' \end{array}. \tag{21}$$

We then have that

$$\mathrm{Tr}_{(S',S)} \mathcal{A} \bullet : L(\mathcal{H}_S) \to L(\mathcal{H}_{S'})$$

$$= \boxed{\mathcal{A}} \bullet \tag{22}$$

$$= \sum_{ij} \langle i_S | \mathcal{A}\Big(\bullet \otimes |i_{S'}\rangle\langle j_{S'}|\Big)|j_S\rangle,$$

where $\{|i_S\rangle\}$ and $\{|i_{S'}\rangle\}$, are orthonormal bases for the two respective systems.

## II. OPEN QUANTUM SYSTEMS

In general, the evolution of an open quantum system is given by a Hamiltonian for the system and a bath, $H(t) = H_S(t) + H_B(t) + V(t)$, together with an initial state $\rho_0$ for the composite system, usually assumed to be in product form $\rho_0 = \rho_S \otimes \rho_B$. Here $H_S(t)$ is a system-Hamiltonian, $H_B(t)$ is the Hamiltonian for the bath, and $V(t)$ is an interaction Hamiltonian. In general, we allow these Hamiltonians to be time-dependent. The reduced state of the system, $\rho_S(t)$, at a time $t$, can then formally be written

$$\rho_S(t) = \mathrm{tr}_B \mathcal{W}(t)\rho_0, \tag{23}$$

where $\mathrm{tr}_B$ denotes a partial trace over the bath, and we have introduced a time-evolution superoperator defined through

$$\mathcal{W}(t)\rho = \mathcal{T} \exp\left(-i \int_0^t \mathrm{d}t' H(t')\right) \rho \mathcal{T} \exp\left(i \int_0^t \mathrm{d}t' H(t')\right). \tag{24}$$

Here $\mathcal{T}$ denotes the time-ordering operator. The method we introduce in the following sections is based on "trotterizing" $\mathcal{W}(t)$. That is, using a Suzuki-Trotter decomposition [3], we seek to write $\mathcal{W}(t)$ as a limit,

$$\mathcal{W}(t) = \lim_{N \to \infty} \mathcal{U}_N \ldots \mathcal{U}_1, \tag{25}$$

where $\mathcal{U}_n \rho = U_n \rho U_n^\dagger$ is a unitary conjugation. For a common class of system-bath Hamiltonians, we show, in a sense that becomes clear below, that $\mathcal{U}_n$ can be made arbitrarily close to the identity map as $N$ is increased. Eq. (25) can thus be understood as a product integral [4] with a well-defined continuum limit as $N \to \infty$.

For finite $N$, the product in Eq. (25) can be represented diagrammatically using the notation introduced in the previous section. This can in certain cases expose a structure to the problem that would not easily be seen otherwise. This is indeed the case for the feedback problem considered in the main letter, and we exploit this to find a product formula for a time-propagator after tracing out the bath degrees of freedom:

$$\mathcal{E}(t) = \lim_{N \to \infty} \mathcal{E}_N \ldots \mathcal{E}_1, \tag{26}$$

where $\mathcal{E}_n$ is a superoperator on an extended system, $S'$, that approaches the identity map as $N \to \infty$. The state $\rho_S(t)$ is finally found by acting with $\mathcal{E}(t)$ on an initial state $\rho_S(0)$ and performing a generalized partial trace, of the type introduced in Eq. (22).

Although this approach is particularly useful for the non-Markovian feedback problem, it is instructive to apply these ideas first to the simpler problem of a Markovian bath, *i.e.*, a conventional reservoir without any feedback loops. This is done in the next section, before we subsequently attack the feedback problem of the main letter.

## III. MARKOVIAN DYNAMICS

We assume that the system's internal dynamics is described by a Hamiltonian, $H_S$, that we leave arbitrary. The coupling to the bath is assumed to be linear (*e.g.*, a dipole coupled to the electric field). More specifically, we assume a system-bath Hamiltonian, $H = H_S + H_B + V$, where (in units where $\hbar = 1$)

$$H_B = \int d\omega \, \omega b^\dagger(\omega) b(\omega) \tag{27}$$

is the bath Hamiltonian, and $V$ is the interaction Hamiltonian, given by

$$V = i \int d\omega \sqrt{\frac{\kappa}{2\pi}} \left( a b^\dagger(\omega) - a^\dagger b(\omega) \right). \tag{28}$$

Here, $\kappa$ is a rate describing the coupling strength of the system to the field, $a$ is an arbitrary system operator, and the field modes, $b(\omega)$, satisfy $[b(\omega), b^\dagger(\omega')] = \delta(\omega - \omega')$.

It is now convenient to go to a rotating frame with respect to the bath Hamiltonian $H_B$. The interaction Hamiltonian then takes the from

$$V(t) = i \int d\omega \sqrt{\frac{\kappa}{2\pi}} \left( a b^\dagger(\omega) e^{i\omega t} - a^\dagger b(\omega) e^{-i\omega t} \right) = i \left( L b^\dagger(t) - L^\dagger b(t) \right), \tag{29}$$

where in the last equality, we have introduced the system operator $L = \sqrt{\kappa} a$, and the time-domain field

$$b(t) = \frac{1}{\sqrt{2\pi}} \int d\omega \, e^{-i\omega t} b(\omega), \tag{30}$$

which satisfies $[b(t), b^\dagger(t')] = \delta(t - t')$. Formally, the solution for the system state at a time $t$ can be written

$$\rho_S(t) = \text{tr}_B \mathcal{W}(t) \rho_S(0), \tag{31}$$

where the evolution superoperator $\mathcal{W}(t)$ is given through

$$\mathcal{W}(t) \rho = \mathcal{T} \exp\left( -i \int_0^t dt' H(t') \right) \rho \bar{\mathcal{T}} \exp\left( i \int_0^t dt' H(t') \right), \tag{32}$$

where $H(t) = H_S + V(t)$ is the Hamiltonian in the rotating frame with respect to $H_B$.

The next step is to go to discretized time. This allows us to expose the structure of the evolution operator in terms of a "tensor network," using the diagrammatic notation from Sect. I. We do this by first introducing ladder operators

$$b_n = \frac{1}{\sqrt{\Delta t}} \int_{t_n - \Delta t}^{t_n} b(s) ds, \tag{33}$$

which satisfy the usual commutation relation $[b_n, b_m^\dagger] = \delta_{nm}$. Here, $t_n = n\Delta t$, for $n = 1, \ldots, N$, where $t = N\Delta t$. Using the Suzuki-Trotter formula [3], the evolution operator can be written

$$\mathcal{W}(t) = \lim_{N \to \infty} \mathcal{U}_N \ldots \mathcal{U}_2 \mathcal{U}_1. \tag{34}$$

Here, and in the following, $N \to \infty$ refers to the continuum limit where $N\Delta t = t$ is held fixed. The unitary superoperator $\mathcal{U}_n$ is defined through $\mathcal{U}_n \bullet = U_n \bullet U_n^\dagger$ with

$$U_n = \exp\left( -i\Delta t H_S + \sqrt{\Delta t} (L b_n^\dagger - L^\dagger b_n) \right). \tag{35}$$

We now work with finite $N$, and eventually take the continuum, $N \to \infty$, limit. Eq. (34) and Eq. (35) then suggest the following picture: The system interacts with a collection of harmonic oscillators, labelled by $n$, where the $n$'th oscillator interacts with the system at time $t_n$, for a short

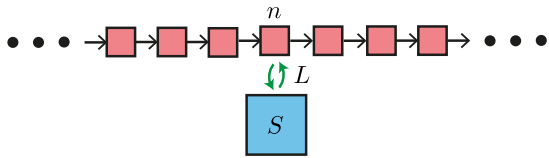

FIG. 1: A "conveyor belt" of harmonic oscillators, labeled by $n$, moves past the system as shown in the figure. The $n$'th oscillator interacts with the system for a short time $\Delta t$.

time $\Delta t$, before leaving the system. After this it never interacts with the system again. We refer to this as the "conveyor belt picture," and it is illustrated pictorially in Fig. 1.

The superoperator $\mathcal{U}_n$ we denote diagramatically as

$$\mathcal{U}_n = \;\; S -\boxed{\mathcal{U}_n}- S \;, \tag{36}$$

where the labels indicate the system associated to the legs: $S$ for the system and $n$ for the $n$'th oscillator. The result of acting with this superoperator on a product state $\rho_S \otimes \sigma_n$ is then denoted

$$\mathcal{U}_n \rho_S \otimes \sigma_n = -\boxed{\mathcal{U}_n}-\boxed{\rho_S}. \tag{37}$$

We are now in a position to represent diagrammatically the system-bath state at time $t_N$, assuming an initial state $\rho_S \otimes \sigma_1 \otimes \sigma_2 \otimes \cdots \otimes \sigma_N$:

$$\rho_{SB}(t_N) = -\boxed{\mathcal{U}_N}- \;\; \cdots \;\; -\boxed{\mathcal{U}_2}-\boxed{\mathcal{U}_1}-\boxed{\rho_S} \;. \tag{38}$$

As noted in the previous section, the shapes appearing in the diagram above, Eq. (38), can be thought of as tensors. A diagram consisting of many tensors, connected by joining their legs, like in Eq. (38), is therefore often called a "tensor network" [5]. In fact, the tensor network representation of the state in Eq. (38) is analogous to a matrix product state (MPS) [6], but with superoperators and density operators in place of operators and kets, respectively. MPS are famous for their role as a variational ansatz in the density matrix renormalization group method [6], used to find the low-energy spectrum of many-body Hamiltonians. In light of this similarity, we call a state on the form of Eq. (38) a *superoperator product state* (SPS) [14]. If taking the continuum limit (more on this below), $N \to \infty$, the resulting state we can similarly call a *continuous* SPS (cSPS), in analogy with the recently introduces cMPS [7]. The connection between output fields of open quantum systems and cMPS has been exploited to use open quantum systems to *generate* cMPS states, in the context of quantum simulation, both in theory [8, 9] and in experiment [10].

In the present context, however, we are chiefly interested in the reduced system dynamics, $\rho_S(t) = \mathrm{tr}_B \rho_{SB}(t)$. That is, we want to trace over the bath degrees of freedom in Eq. (38). This gives a reduced system state (cf. Eq. (14)):

$$\rho_S(t_N) = -\boxed{\mathcal{U}_N}- \;\; \cdots \;\; -\boxed{\mathcal{U}_2}-\boxed{\mathcal{U}_1}-\boxed{\rho_S} \;. \tag{39}$$

We would next like to retrieve the continuum, $N \to \infty$, limit, to find a more compact form of Eq. (31) for $\rho_S(t)$. To keep the discussion as simple as possible, we assume an incoming vacuum

field, *i.e.*, we assume $\sigma_n = |0\rangle\langle 0|$, for all $n$. We can then find an analytical expression for the map $\mathcal{E}_n$,

$$\mathcal{E}_n(\bullet) = \boxed{\mathcal{U}_n}\,\bullet\, = \text{tr}_n \mathcal{U}_n(\bullet \otimes \sigma_n) \tag{40}$$

$$= \exp(-i\Delta t \mathcal{H}[H_S]\bullet + \Delta t \mathcal{D}[L]\bullet), \tag{41}$$

where we have defined superoperators

$$\mathcal{H}[X]\bullet = [X, \bullet], \tag{42}$$

$$\mathcal{D}[X]\bullet = X\bullet X^\dagger - \frac{1}{2}X^\dagger X \bullet - \frac{1}{2}\bullet X^\dagger X. \tag{43}$$

In Eq. (40) higher order terms in $\Delta t$ have been neglected. In the continuum limit, $N \to \infty$, we define a time-propagator, $\mathcal{E}(t)$, which can be written as a product integral [4]:

$$\mathcal{E}(t) = \lim_{N\to\infty} \prod_{n=1}^{N} \mathcal{E}_n. \tag{44}$$

For the case of a vacuum bath, this is simply

$$\mathcal{E}(t) = \exp(-it\mathcal{H}[H_S] + t\mathcal{D}[L]). \tag{45}$$

The Lindblad equation is retrieved by differentiating:

$$\frac{\mathrm{d}}{\mathrm{d}t}\mathcal{E}(t) = -i\mathcal{H}[H_S]\mathcal{E}(t) + \mathcal{D}[L]\mathcal{E}(t). \tag{46}$$

The initial condition for the time-propagator is $\mathcal{E}(0) = \mathcal{I}_S$, where $\mathcal{I}_S = I_S \bullet I_S$ is the system identity superoperator. Finally, the system state, $\rho_S(t)$, is found through

$$\rho_S(t) = \mathcal{E}(t)\rho_S(0). \tag{47}$$

## IV. A SOLUTION TO THE FEEDBACK PROBLEM

### A. A diagrammatic representation in discretized time

Equipped with the graphical notation, and having warmed up with a Markovian reservoir in the previous section, we are ready to attack the feedback problem of the main letter.

Recall that we are now considering the following interaction Hamiltonian:

$$\begin{aligned}V = & i\int_{-\infty}^{\infty} \mathrm{d}\omega \sqrt{\frac{\kappa_1}{2\pi}}\left(a_1 b^\dagger(\omega) - \text{H.c.}\right) \\ & + i\int_{-\infty}^{\infty} \mathrm{d}\omega \sqrt{\frac{\kappa_2}{2\pi}}\left(a_2 b^\dagger(\omega)e^{-i\omega\tau + i\phi} - \text{H.c.}\right),\end{aligned} \tag{48}$$

where $\tau = l/c$ is the time-delay, $\sqrt{\kappa_{1,2}}$ is the coupling strength at the two positions, $x = 0, l$, respectively, and $a_1$ and $a_2$ are two system operators. We follow the steps of the previous section, and go to a rotating frame with respect to the bath Hamiltonian $H_B = \int \mathrm{d}\omega \omega b^\dagger(\omega) b(\omega)$:

$$\begin{aligned}V(t) = & i\int \mathrm{d}\omega \sqrt{\frac{\kappa_1}{2\pi}}\left(a_1 b^\dagger(\omega)e^{i\omega t} - \text{H.c.}\right) \\ & + i\int \mathrm{d}\omega \sqrt{\frac{\kappa_2}{2\pi}}\left(a_2 b^\dagger(\omega)e^{i\omega(t-\tau)+i\phi} - \text{H.c.}\right)\end{aligned} \tag{49}$$

$$= i\left(L_1 b^\dagger(t) - L_1^\dagger b(t)\right) + i\left(L_2 b^\dagger(t-\tau) - L_2^\dagger b(t-\tau)\right), \tag{50}$$

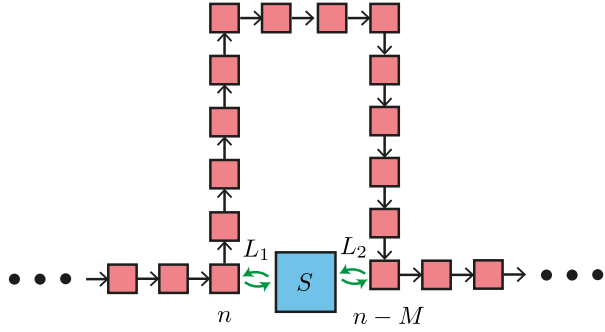

FIG. 2: A "conveyor belt" of harmonic oscillators, labeled by $n$, moves past the system. The $n$'th and the $(n-M)$'th oscillator interacts with the system simultaneously, for a short time, $\Delta t$.

where we have introduced the time-domain field, $b(t)$, defined as in Eq. (30), and $L_1 = \sqrt{\kappa_1}a_1$, $L_2 = e^{i\phi}\sqrt{\kappa_2}a_2$. The evolution superoperator, $\mathcal{W}(t)$ defined as before in Eq. (32), can again be written in terms of a product integral using the Suzuki-Trotter formula:

$$\mathcal{W}(t) = \lim_{N \to \infty} \mathcal{U}_N \mathcal{V}_{N-M} \ldots \mathcal{U}_2 \mathcal{V}_{2-M} \mathcal{U}_1 \mathcal{V}_{1-M}. \tag{51}$$

Here $\mathcal{U}_n = U_n \bullet U_n^\dagger$ and $\mathcal{V}_n = V_n \bullet V_n^\dagger$ are unitary conjugations with

$$\begin{aligned} U_n &= \exp\left(-i\frac{\Delta t}{2} H_S + i\sqrt{\Delta t}(L_1 b_n^\dagger - L_1^\dagger b_n)\right), \\ V_n &= \exp\left(-i\frac{\Delta t}{2} H_S + i\sqrt{\Delta t}(L_2 b_n^\dagger - L_2^\dagger b_n)\right), \end{aligned} \tag{52}$$

where $b_n$ is defined as in Eq. (33). $N\Delta t = t$ and $M\Delta t = \tau$ are both to be held fixed as $N \to \infty$.

Eq. (51) and Eq. (52) suggest the following picture: The system interacts with a collection of harmonic oscillators, labelled by $n$, where the $n$'th oscillator interacts with the system at time $t_n$, for a short time $\Delta t$, before leaving the system. The *same* oscillator comes back to interact with the system a time $\tau = M\Delta t$ later. After this it never interacts with the system again. This is illustrated pictorially in Fig. 2.

We assume that any oscillator that has not yet interacted with the system is in a product state with everything else. We also assume that at time $t = 0$, the initial state is on the form $\rho_S \otimes \sigma_{1-M} \otimes \cdots \otimes \sigma_1 \otimes \cdots \otimes \sigma_N$. The evolution up to time $\tau = t_M = M\Delta t$ is then essentially entirely analogous to what we had in the previous section. Diagrammatically, we can write

$$\rho_{SB}(t_M) = \ \text{[diagram: } \mathcal{U}_M\text{-}\mathcal{V}_0 \cdots \mathcal{U}_1\text{-}\mathcal{V}_{1-M}\text{-}\rho_S\text{]}, \tag{53}$$

where we have suppressed the labels on the oscillator states, $\sigma_n$, for notational convenience. The oscillators with $n < 1$ will at this point never interact with the system again, and can therefore be traced out. Diagramatically this is represented by

$$\rho_{SL}(t_M) = \ \text{[diagram: } \mathcal{U}_M\text{-}\mathcal{V}_0 \cdots \mathcal{U}_1\text{-}\mathcal{V}_{1-M}\text{-}\rho_S\text{]}. \tag{54}$$

Here $L$ stands for "loop," and refers to the subset of bath-oscillators that have interacted with the system exactly once. Now, to integrate to time $t_{M+1}$ we need to attach the two superoperators $\mathcal{V}_1$

and $\mathcal{U}_{M+1}$ to Eq. (54):

$$\rho_{SL}(t_{M+1}) = \boxed{\mathcal{U}_{M+1}\, \mathcal{V}_1\, \mathcal{U}_M\, \mathcal{V}_0\, \cdots\, \mathcal{U}_1\, \mathcal{V}_{1-M}\, \rho_S}. \tag{55}$$

Notice that since $\mathcal{V}_1$ acts on the same oscillator as $\mathcal{U}_1$, the upper (input) leg of the former has to be connected to the lower (output) leg of the latter. We can equivalently redraw this diagram in the following way:

$$\rho_{SL}(t_{M+1}) = \boxed{\mathcal{U}_M\, \mathcal{V}_0\, \cdots\, \mathcal{U}_1\, \mathcal{V}_{1-M}\, \rho_S \atop \mathcal{U}_{M+1}\, \mathcal{V}_1}. \tag{56}$$

It is stressed that Eq. (55) and Eq. (56) are equivalent diagrams, drawn in two different ways. The latter, Eq. (56), is the preferred way of drawing the diagram, as will become clear in the following. In fact, this is the crux of the problem, and here the power of the diagrammatic notation comes into play. The diagrammatic notation is equivalent to the usual algebraic Dirac notation, but the same expression can be drawn in several ways, and the diagrammatic notation can therefore be richer than the algebraic notation. In some situations, as is the case here, the diagrammatic notation can reveal a structure to the problem that would not easily be seen otherwise.

As we continue evolving the state for times $\tau < t_N \leq 2\tau$, we keep attaching alternating $\mathcal{V}$'s and $\mathcal{U}$'s to the second row in Eq. (56). So, for example, the state at time $t_{2M} = 2\tau$, we draw

$$\rho_{SL}(t_{2M}) = \boxed{\mathcal{U}_M\, \mathcal{V}_0\, \cdots\, \mathcal{U}_1\, \mathcal{V}_{1-M}\, \rho_S \atop \mathcal{U}_{2M}\, \mathcal{V}_M\, \cdots\, \mathcal{U}_{M+1}\, \mathcal{V}_1}. \tag{57}$$

At this point we can do the same trick as we did at time $t_M$, and start a new row to attach the next pair of superoperators. That is, the state at time $t_{2M+1}$ we can draw

$$\rho_{SL}(t_{2M+1}) = \boxed{\mathcal{U}_M\, \mathcal{V}_0\, \cdots\, \mathcal{U}_1\, \mathcal{V}_{1-M}\, \rho_S \atop \mathcal{U}_{2M}\, \mathcal{V}_M\, \cdots\, \mathcal{U}_{M+1}\, \mathcal{V}_1 \atop \mathcal{U}_{2M+1}\, \mathcal{V}_{M+1}}. \tag{58}$$

And so it continues: We evolve the state by attaching $\mathcal{V}$'s and $\mathcal{U}$'s, and start a new row at each $n = kM + 1$, where $k = 1, 2, \ldots$. Now, the system state $\rho_S(t_N) = \mathrm{tr}_L \rho_{SL}(t_N)$ at time $t_N$, is found

by tracing over the field. For example, the system state at time $t_{2M+1}$ is found from Eq. (58):

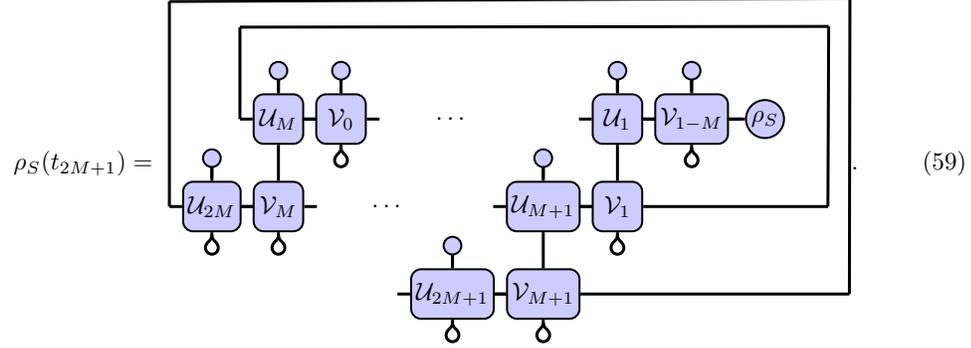

$$\rho_S(t_{2M+1}) = \quad . \tag{59}$$

Notice that there is only one unpaired leg, going to the left, as it should be for a reduced system state.

We are now ready to represent diagramatically a formal solution for the system state at an arbitrary time, $(k-1)\tau \le t_N < k\tau$, where $k = 1, 2, \dots$:

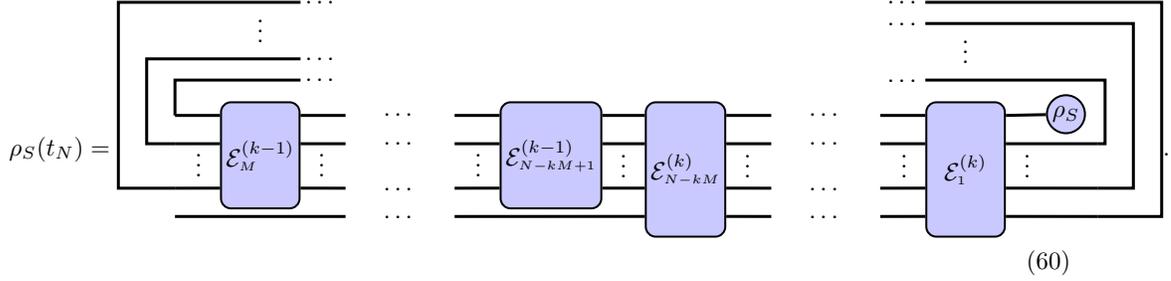

$$\rho_S(t_N) = \quad . \tag{60}$$

Here we have introduced the tensor $\mathcal{E}_n^{(k)}$:

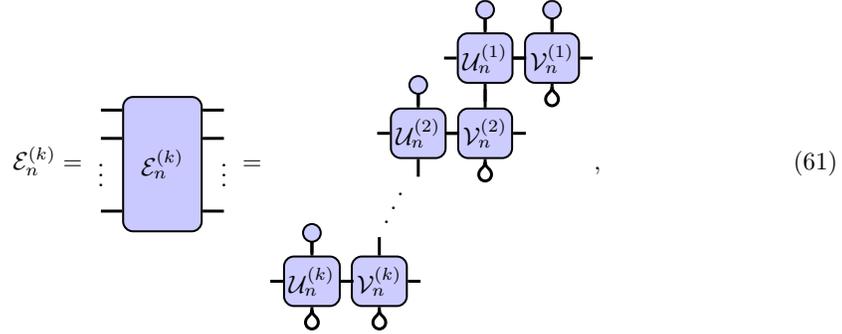

$$\mathcal{E}_n^{(k)} = \quad = \quad , \tag{61}$$

where $\mathcal{U}_n^{(l)}$ is shorthand for $\mathcal{U}_{n+(l-1)M}$ and $\mathcal{V}_n^{(l)}$ for $\mathcal{V}_{n+(l-2)M}$.

We can think of $\mathcal{E}_n^{(k)}$ as a superoperator acting on $k$ copies of the system which we can label $S_l$, for $l = 1, \dots, k$. A compact way to write $\rho_S(t_N)$ can be found by introducing a time-propagator on a $k$-fold system space, $\mathcal{E}_M(t_N)$:

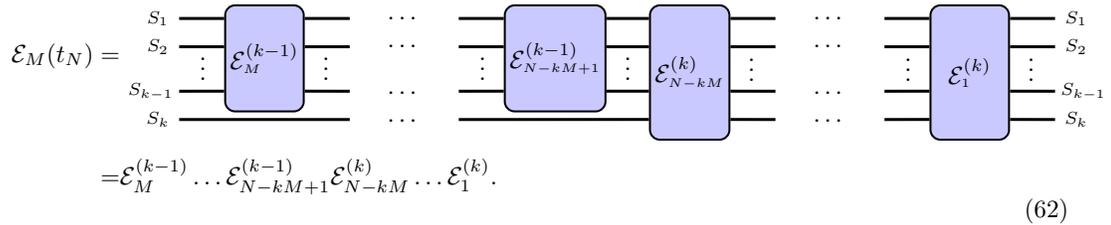

$$\mathcal{E}_M(t_N) =$$

$$= \mathcal{E}_M^{(k-1)} \dots \mathcal{E}_{N-kM+1}^{(k-1)} \mathcal{E}_{N-kM}^{(k)} \dots \mathcal{E}_1^{(k)}. \tag{62}$$

$\rho_S(t_N)$ is found by imposing the following boundary conditions:

$$\rho_S(t_N) = \text{Tr}_{(S_k,S_{k-1})} \ldots \text{Tr}_{(S_2,S_1)} \mathcal{E}_M(t_N) \rho_{S_1}, \tag{63}$$

where the generalized trace, Tr, is defined in Eq. (22), and $\rho_{S_1}$ is the system initial state.

### B. The continuum limit

We wish to take the continuum limit of Eq. (62). Again, we assume for simplicity an incoming vacuum field, with $\sigma_n = |0\rangle\langle 0|$ for all $n$. We first note that $\mathcal{E}_n^{(k)}$ can be built from the following tensors:

$$\mathcal{T}_n^{(k)}(\bullet) = \boxed{\mathcal{U}_n^{(k)}} \bullet = \text{tr}_{n(k)} \mathcal{U}_n^{(k)} \left( \bullet \otimes \sigma_{n(k)} \right) \tag{64}$$

$$= \exp\left(-i\frac{\Delta t}{2} \mathcal{H}[H_S^{(k)}] \bullet + \Delta t \mathcal{D}[L_1^{(k)}] \bullet\right), \tag{65}$$

where we have defined $n(l) = n + (l-1)M$ and the superscript denotes the system on which an operator acts,

$$\mathcal{T}_n^{(1)}(\bullet) = \boxed{\mathcal{V}_n^{(1)}} \bullet = \text{tr}_{n(0)} \mathcal{V}_n^{(1)} \left( \bullet \otimes \sigma_{n(0)} \right) \tag{66}$$

$$= \exp\left(-i\frac{\Delta t}{2} \mathcal{H}[H_S^{(1)}] \bullet + \Delta t \mathcal{D}[L_2^{(1)}] \bullet\right), \tag{67}$$

and the composite tensor

$$\mathcal{T}_n^{(l+1,l)}(\bullet) = \boxed{\begin{array}{c}\mathcal{U}_n^{(l)}\\ \mathcal{V}_n^{(l+1)}\end{array}} \bullet = \text{tr}_{n(l)} \mathcal{V}_{n(l)}^{(l+1)} \mathcal{U}_{n(l)}^{(l)} \left( \bullet \otimes \sigma_{n(l)} \right) \tag{68}$$

$$= \exp\left(-i\frac{\Delta t}{2} \mathcal{H}[H_S^{(l+1)} + H_S^{(l)}] \bullet + \Delta t \mathcal{D}[L_1^{(l)}] \bullet + \Delta t \mathcal{D}[L_2^{(l+1)}] \bullet + \Delta t \mathcal{C}[L_1^{(l)}, L_2^{(l+1)}] \bullet\right), \tag{69}$$

where

$$\mathcal{C}[X,Y]\bullet = X \bullet Y^\dagger + Y \bullet X^\dagger - Y^\dagger X \bullet - \bullet Y^\dagger X. \tag{70}$$

$\mathcal{E}_n^{(k)}$ can be written in terms of these tensors as a product

$$\begin{aligned}\mathcal{E}_n^{(k)} &= \mathcal{T}_n^{(k)} \mathcal{T}_n^{(k,k-1)} \ldots \mathcal{T}_n^{(3,2)} \mathcal{T}_n^{(2,1)} \mathcal{T}_n^{(1)}\\ &= \exp\left(\sum_{l=0}^k -\frac{i\Delta t}{2} \mathcal{H}\left[H_{l,l+1}\right] + \Delta t \mathcal{D}\left[L_{l,l+1}\right]\right),\end{aligned} \tag{71}$$

where, for notational convenience, we have defined

$$H_{l,l+1} = H_S^{(l)} + H_S^{(l+1)} + i(L_1^{(l)\dagger} L_2^{(l+1)} - \text{H.c.}), \tag{72}$$

$$L_{l,l+1} = L_1^{(l)} + L_2^{(l+1)}, \tag{73}$$

except for

$$H_{0,1} = H_S^{(1)}, \tag{74}$$

$$H_{k,k+1} = H_S^{(k)}, \tag{75}$$

$$L_{0,1} = L_2^{(1)}, \tag{76}$$

$$L_{k,k+1} = L_1^{(k)}. \tag{77}$$

We can now write the continuum limit of Eq. (62) as

$$\begin{aligned}\mathcal{E}_\tau(t) &= \lim_{N \to \infty} \mathcal{E}_M^{(k-1)} \ldots \mathcal{E}_{N-kM+1}^{(k-1)} \mathcal{E}_{N-kM}^{(k)} \ldots \mathcal{E}_1^{(k)} \\ &= \mathcal{T} \exp\left(\int_0^\tau ds \sum_{l=0}^{k} \left\{ -\frac{i}{2}\mathcal{H}[H_S^{(l+1)}(s) + H_S^{(l)}(s)] + \mathcal{D}[L_{l,l+1}(s)] \right\}\right),\end{aligned} \tag{78}$$

where $\mathcal{T}$ is the time-ordering operator, and we have defined

$$A^{(l)}(s) = A^{(l)} \quad \text{for all } l \neq k, \tag{79}$$

$$A^{(k)}(s) = \theta[t - (k-1)\tau - s] A^{(k)}, \tag{80}$$

for any system operator $A$, where $\theta(t)$ is the Heaviside step function.

We can differentiate Eq. (78) to find a differential equation for $\mathcal{E}_s(t)$:

$$\frac{d}{ds}\mathcal{E}_s(t) = \sum_{l=0}^{k} \left\{ -\frac{i}{2}\mathcal{H}\left[H_{l,l+1}(s)\right] + \mathcal{D}\left[L_{l,l+1}(s)\right] \right\} \mathcal{E}_s(t). \tag{81}$$

with initial condition $\mathcal{E}_0(t) \equiv \mathcal{I}_S^{\otimes(k)}$.

The continuum solution for $\rho_S(t)$ is found from $\mathcal{E}_\tau(t)$ by imposing the boundary conditions in Eq. (63):

$$\rho_S(t) = \text{Tr}_{(S_k, S_{k-1})} \ldots \text{Tr}_{(S_2, S_1)} \mathcal{E}_\tau(t) \rho_{S_1}, \tag{82}$$

where $\rho_{S_1}$ is the initial state. Eq. (81) and Eq. (82) are the main results stated in the letter.

[1] R. Penrose, Comb. Math. Appl (1971).
[2] C. J. Wood, J. D. Biamonte, and D. G. Cory, Quant. Inf. Comp. **15**, 0579 (2015).
[3] M. Suzuki, Comm. Math. Phys. **51**, 183 (1976).
[4] J. D. Dollard and C. N. Friedman, J. Funct. Anal. **28**, 309 (1978).
[5] R. Orús, Ann. Phys. **349**, 117 (2014).
[6] U. Schollwöck, Ann. Phys. **326**, 96 (2011).
[7] F. Verstraete and J. I. Cirac, Phys. Rev. Lett. **104**, 190405 (2010).
[8] T. J. Osborne, J. Eisert, and F. Verstraete, Phys. Rev. Lett. **105**, 260401 (2010).
[9] S. Barrett, K. Hammerer, S. Harrison, T. Northup, and T. Osborne, Phys. Rev. Lett. **110**, 090501 (2013).
[10] C. Eichler et al., in preparation (2015).
[11] J. E. Gough, M. R. James, and H. I. Nurdin, New J. Phys. **16**, 075008 (2014).
[12] Sometimes the "vectorization" of an operator $\tau$ is denoted by a "ket", $|\tau\rangle\rangle$.
[13] When working in Liouville space, the legs can be thought of as representing "double indices." For example, if the system dimension is $d$, this would be an index running over $d^2$ values.
[14] In other contexts the term MPS is usually used to refer to a *pure* state of the bath degrees of freedom only, *e.g.*, after projecting onto a particular system state. Restricting to pure bath states implies a restriction on the unitaries, $\mathcal{U}_n$. What we here call "the bath" is then a many-body system of interest, and what we refer to as "the system" only plays an ancillary role. In the context of open quantum systems, however, what we here call the system is the central object of interest, and it is therefore natural to refer to a system-bath state such as Eq. (38) as an SPS. This is in line with the terminology used recently in Ref. [11] for continuous matrix-product states (cMPS) arising in the context of open systems.



\end{document}